\documentstyle[aps,prb,psfig]{revtex}

\begin{document}

\title{Nonadiabatic pairing effects for tight-binding
electrons interacting with phonons} 
\author{A. Perali$^{1}$, 
C. Grimaldi$^{1,2}$ and L. Pietronero$^{1,2}$}

\address{$^{1}$Dipartimento di Fisica, Universit\'{a} di Roma ``La
Sapienza", 
Piazzale A.  Moro, 2, 00185 Roma, Italy }
\address{$^{2}$ Istituto Nazionale Fisica della Materia, Unit\'a di Roma 1, Italy}
\date{\today} 
\maketitle 
\medskip

\begin{abstract}
The nonadiabatic electron-phonon corrections for the superconducting 
pairing are investigated for a specific tight-binding model
corresponding to a $2d$ square lattice.
This permits to investigate the role of various specific
properties like the band filling, nesting effects and a realistic
van Hove singularity on the superconducting effective pairing beyond
Migdal's limit.
The main results are: $(i)$ Starting from a momentum independent
electron-phonon coupling the nonadiabatic effects lead to an
effective pairing which is strongly dependent on frequency and momentum.
$(ii)$ If instead the electron-phonon coupling is
mainly forward (as due to correlation effects) the resulting
pairing results to be strongly enhanced.
These results confirm but also extend the simplified scheme
used up to now to compute these properties. In this respect our
results link the nonadiabatic effects to the specific properties
of realistic materials.
\end{abstract}
\vskip 2pc 

\section{Introduction}
\label{intro}
Unconventional superconductors, such as cuprates and fullerides,
are characterized by Fermi energies $E_F$ much smaller than those
of conventional metals.\cite{uemura,plakida,gunna} 
In this situation, the energy scale
$\omega_0$ of the mediator of the superconducting pairing can be
comparable to $E_F$ so that the quantity $\omega_0/E_F$ is no longer
negligible as it happens for conventional 
superconductors.
As a consequence, vertex corrections in the normal state electronic
self-energy become relevant because of the non-validity of Migdal's
approximation \cite{migdal} and also the Migdal-Eliashberg (ME) approach
to the superconducting state\cite{migdal,elia} is 
drastically modified.\cite{schrieffer}

In recent years we have generalized both the superconducting and
the normal states in order to include the first corrections beyond Migdal's 
limit.\cite{GPSprl,PSG,CP,GCP,CGP} 
We considered the energy scale $\omega_0$ to be given by phonons so that
the parameter $\omega_0/E_F$ measures the degree of adiabaticity
of the lattice dynamics compared to the electronic kinetic energy.
For $\omega_0/E_F\ll 1$ and electron-phonon ($e$-$ph$) 
coupling  
$\lambda <1$ one recovers the ME theory of the
electron-phonon coupled system. By enhancing the coupling $\lambda$
the electron develops a large effective mass (polaron) and eventually
forms bi-polaronic states. However, the $e$-$ph$ coupled system
may display a regime which is different from both the ME picture
and the polaronic (bi-polaronic) regime. We have named this situation 
{\it nonadiabatic}, and it is characterized by quasi-free electrons,
{\it i.e.}, $\lambda <1$, coupled in a nonadiabatic way ($\omega_0/E_F$
not negligible) to the lattice. We have shown that in this 
regime it is possible to
define a perturbative theory where the expansion parameter is roughly
given by $\lambda\omega_0/E_F$. 
Moreover, the corrections arising from the nonadiabatic
hypothesis have a non-trivial structure in both the momentum transfer $q$
and the Matsubara exchanged frequency $\omega$. 
In fact, the vertex correction appearing in the normal state self-energy 
becomes positive (negative) when $v_F q<\omega$ ($v_F q>\omega$),
where $v_F$ is the Fermi velocity.\cite{PSG}
The generalization to the superconducting
transition reveals that this situation is also encountered for the class
of diagrams beyond Migdal's limit relevant for the Cooper channel.
Concerning the critical temperature $T_c$,  as long as
the momentum transfer is less than $\omega_0/v_F$, the nonadiabatic
corrections lead to a strong enhancement of $T_c$ also for moderate
values of the $e$-$ph$ coupling $\lambda$.\cite{GPSprl,PSG,CP}
The opening of a superconducting gap changes drastically the
behavior in momentum-frequency space of the vertex especially
in the low temperature limit. In this regime the vertex can be
positive also when $v_F q>\omega$ probably leading to
anomalous values of the gap-to-$T_c$ ratio.\cite{CGP}

The results presented in the previous papers on the nonadiabatic
$e$-$ph$ interaction theory have been derived by
employing higly simplified approximations. Although several results were
approximation-independent, it is important to study the behavior
of the nonadiabatic corrections by using realistic models.
The aim of this paper is to provide
a study of both the vertex and cross diagrams in the framework
of a specific tight-binding model for the electronic band. 
The advantage
of such a study is twofold: it permits to confirm the results obtained
by using simple models but also the whole picture is made
much richer so that it is possible to relate
specific nonadiabatic properties to specific features of
the material. Finally, this paper can be also regarded as a
preliminary work for a completely numerical solution of
the Eliashberg equations generalized beyond Migdal's limit.

The paper is organized as follows. In Sec.\ref{bme} we introduce the
equations for the superconducting instability driven by the
$e$-$ph$ coupling generalized in order
to include the first nonadiabatic corrections represented by the
vertex and cross diagrams for a specific tight-binding model. 
In Sec.\ref{indep} we study the vertex
and cross corrections for a model in which the $e$-$ph$ interaction
is momentum independent while in Sec.\ref{smallq} we analyze the
effect on the nonadiabatic corrections of an $e$-$ph$ coupling
peaked at small values of the momentum transfer as induced by electronic
correlations. The last section is devoted to the conclusions.

\section{Beyond Migdal-Eliashberg theory}
\label{bme}
\begin{figure}
\protect
\centerline{\psfig{figure=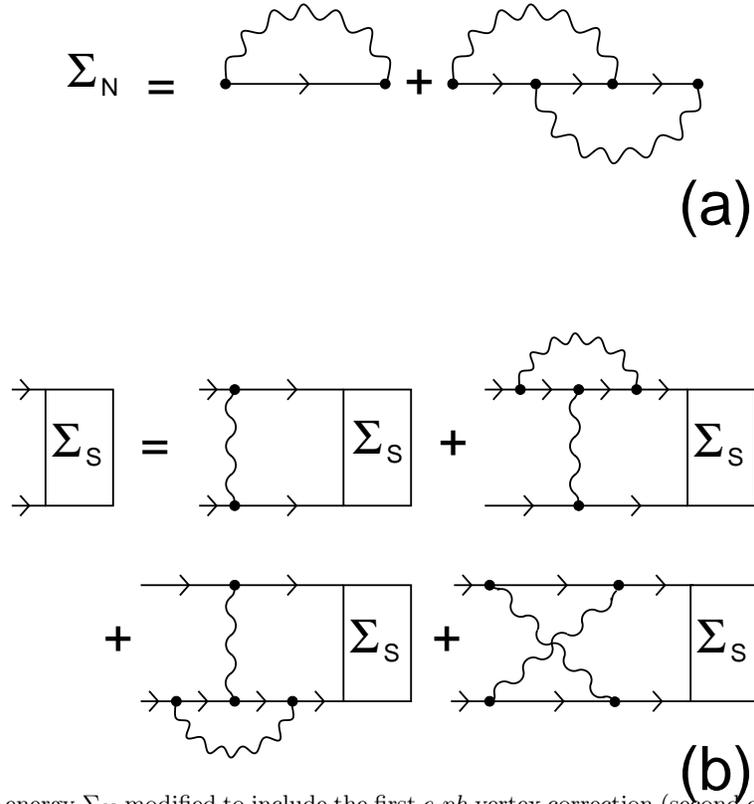,width=10cm}}
\caption{(a): electron self-energy $\Sigma_N$ modified to include
the first $e$-$ph$ vertex correction (second diagram).
(b): self-consistent equation for the off-diagonal 
electron self-energy. The ME theory retains only the first
contribution while the first nonadiabatic corrections add
the vertex (second and third diagrams) and the cross (last diagram)
terms} 
\label{diagrams}
\end{figure}

The presence of nonadiabatic corrections beyond Migdal's
limit affects both the normal and superconducting properties of the
metallic systems in a non-trivial way.
In fact, as it has been already shown in Refs.\onlinecite{GPSprl,PSG},
up to the first corrections of order $\lambda\omega_0/E_F$, the
normal state electronic self-energy is modified by only one nonadiabatic
correction, the vertex correction, while for the superconductive instability 
one needs to include also the cross correction. This situation is shown
in Fig.\ref{diagrams} where the normal state electronic 
self-energy $\Sigma_N$
(Fig.\ref{diagrams}a) and the linearized gap equation (Fig.\ref{diagrams}b) 
are depicted in order to include the first corrections beyond Migdal's limit. 

In order to find the critical temperature $T_c$, one has to solve the following coupled set of equations relating the diagonal, $\Sigma_N(k)$, and the 
off-diagonal, $\Sigma_S(k)$, electron self-energies:

\begin{eqnarray}
\label{self1}
\Sigma_N(k) & = & \sum_{k'}V_0(k-k')[1+P(k,k')]G(k'), \\
\label{gap1}
\Sigma_S(k) & = & \sum_{k'}V(k,k')G(k')G(-k')\Sigma_S(k'),
\end{eqnarray}
where $k= (i\omega_k,{\bf k})$, $k'= (i\omega_{k'},{\bf k}')$,
and we use the notation $\sum_k= -T_c\sum_{\omega_k}\sum_{{\bf k}}$.
In the above equations, $V_0(k-k')=g({\bf k}-{\bf k}')^2 D(k-k')$ where
$g({\bf k}-{\bf k}')$ is the $e$-$ph$ matrix element, $G(k)$ and $D(q)$
are the electron and phonon propagators given by:

\begin{equation}
\label{green1}
G^{-1}(k)=i\omega_k-\epsilon({\bf k})-\Sigma_N(k),
\end{equation}
and

\begin{equation}
\label{green2}
D(q)=-\frac{\omega_0^2}{\omega_q^2+\omega_0^2},
\end{equation}
respectively.
In equations (\ref{green1}) and (\ref{green2}), $\epsilon({\bf k})$
is the electronic band dispersion and $\omega_0$ is the phonon frequency
which we assume dispersionless for simplicity.
In Eq.(\ref{self1}), $P(k,k')$ is the first vertex correction beyond Migdal's limit,
which according to Fig.\ref{diagrams}a is given by:

\begin{equation}
\label{ver1}
P(k,k')=\sum_{p}V_0(k-p)G(p-k+k')G(p).
\end{equation}
The total pairing interaction $V(k,k')$
appearing in the linearized gap equation (\ref{gap1})
is instead given by:

\begin{equation}
\label{kernel1}
V(k,k')=V_0(k-k')[1+2P(k,k')]+C(k,k'),
\end{equation}
where $P(k,k')$ is again the vertex correction (\ref{ver1}) while
$C(k,k')$ is the cross term (last diagram appearing in Fig.\ref{diagrams}b):

\begin{equation}
\label{cross1}
C(k,k')=\sum_p V_0(k-p)V_0(p-k')G(p)G(p-k-k').
\end{equation}
The ordinary Migdal-Eliashberg equations are obtained by neglecting
both the vertex $P$ and cross $C$ corrections in 
Eqs.(\ref{self1},\ref{gap1},\ref{kernel1}),
and in this limit the pairing interaction (\ref{kernel1}) 
reduces to $V_0(k-k')$.

From the above equations, it is clear that to obtain the diagonal
and off-diagonal self-energies one has to calculate
the vertex and cross correction $P(k,k')$ and $C(k,k')$. In previous
papers we have presented the first detailed evaluations of these corrections
both for 3-d \cite{GPSprl,PSG} and 2-d systems.\cite{CP} 
However in calculating $P$ and $C$
we have employed several approximations:  the most
important ones are probably the linearization of the electronic 
dispersion and the assumption of half filling. Although the former
approximation is valid for small momentum transfer in the $e$-$ph$
scattering, it becomes poor for momentum transfer of order $k_F$.
The previous calculations moreover did not address the problem of the role of
band filling, which we know to be important for the nonadiabatic corrections.
In fact, it has been shown that the hole-particle contribution present
in the vertex correction is due to the exchange part of the phonon-mediated
electron-electron interaction which strongly depends on the electron 
density.\cite{GPM}

In this paper we present results obtained by numerical integration
of Eqs(\ref{ver1},\ref{cross1}) over the Brilloiun zone by using
a two dimensional electronic dispersion $\epsilon({\bf k})$ with nearest 
and next-nearest neighbors hopping elements ($t,t'$):

\begin{equation}
\label{disp}
\epsilon({\bf k})=
-2t\left[\cos(k_x)+\cos(k_y)\right]+4t'\cos(k_x)\cos(k_y)-\mu,
\end{equation}
where $\mu$ is the chemical potential.
The above electronic dispersion is quite general for $2D$ transport in
strongly correlated systems and it is also suitable to describe the
conduction band associated with the $CuO_2$ planes in the high-$T_c$
superconductors.\cite{Norman}
 
The calculation of the vertex and cross corrections for each $k$ and $k'$
provides  the effective pairing interaction $V(k,k')$ appearing in the 
linearized gap equation (\ref{gap1}).
We show that $V(k,k')$ has a non-trivial momentum and frequency 
dependence which can be in favour of a strongly anisotropic $s$-wave solution
of the gap or a $d$-wave symmetry when a repulsive potential
is added to the kernel. Our calculations are also aimed to show that
the anisotropy provided by the nonadiabatic corrections is an important
element for a possible enhancement of the critical temperature $T_c$.
For this reason, any approximation scheme which neglects such a
momentum-frequency interplay easily leads to an underestimation of the 
nonadiabatic contributions to enhancing $T_c$.

In all the calculations presented below, we have neglected the self-energy contribution
appearing in Eqs.(\ref{ver1},\ref{cross1}) by inserting in the diagrams the
bare propagators $G_0^{-1}(k)=i\omega_k-\epsilon({\bf k})$.
In this way, the summation over
the Matsubara frequencies can be preformed exactly by using 
the analytical continuation and contour integrals in the complex $z$-plane
(Poisson's formula). 
After this summation is carried out, the vertex and the cross
corrections reduces to: 

\begin{eqnarray}
\label{ver2}
& & P(k,k')=P({\bf k},i\omega_k;{\bf k}',i\omega_{k'}) = 
\frac{\omega_0}{2}\frac{1}{N}\sum_{{\bf p}}\frac{g({\bf k}-{\bf p})^2}
{\epsilon({\bf p})-\epsilon({\bf p}-{\bf k}+{\bf k}')
-i(\omega_k-\omega_{k'})} \nonumber \\
& & \times\left[ \frac{f(\epsilon({\bf p}))+n(-\omega_0)}
{\epsilon({\bf p})+\omega_0-i\omega_k}
-\frac{f(\epsilon({\bf p}))+n(\omega_0)}
{\epsilon({\bf p})-\omega_0-i\omega_k} 
- \frac{f(\epsilon({\bf p}-{\bf k}+{\bf k}'))+n(-\omega_0)}
{\epsilon({\bf p}-{\bf k}+{\bf k}')+\omega_0-i\omega_{k'}}+
\frac{f(\epsilon({\bf p}-{\bf k}+{\bf k}'))+n(\omega_0)}
{\epsilon({\bf p}-{\bf k}+{\bf k}')-\omega_0-i\omega_{k'}}
\right] . \nonumber \\
\end{eqnarray} 
and
\begin{eqnarray}
\label{cross2}
& & C(k,k')= C({\bf k},i\omega_k;{\bf k}',i\omega_{k'}) = \frac{\omega_0}{2}
\frac{2\omega_0+i(\omega_k+\omega_{k'})}
{(\omega_k-\omega_{k'})^2+(2\omega_0)^2}
\frac{1}{N}\sum_{{\bf p}}\frac{g({\bf k}-{\bf p})^2 g({\bf p}-{\bf k}')^2}
{\epsilon({\bf p})-\epsilon({\bf p}-{\bf k}-{\bf k}')
-i(\omega_k+\omega_{k'})} \nonumber \\
& & \times\left[\frac{f(\epsilon({\bf p}))+n(\omega_0)}
{(\epsilon({\bf p})-\omega_0-i\omega_k)(\epsilon({\bf p})-\omega_0-i\omega_{k'})}
-\frac{f(\epsilon({\bf p}))+n(-\omega_0)}
{(\epsilon({\bf p})+\omega_0-i\omega_k)(\epsilon({\bf p})+\omega_0-i\omega_{k'})}
\right.\nonumber \\
& &\left. -\frac{f(\epsilon({\bf p}-{\bf k}-{\bf k}'))+n(\omega_0)}
{(\epsilon({\bf p}-{\bf k}-{\bf k}')-\omega_0+i\omega_k)
(\epsilon({\bf p}-{\bf k}-{\bf k}')-\omega_0+i\omega_{k'})}
+\frac{f(\epsilon({\bf p}-{\bf k}-{\bf k}'))+n(-\omega_0)}
{(\epsilon({\bf p}-{\bf k}-{\bf k}')+\omega_0+i\omega_k)
(\epsilon({\bf p}-{\bf k}-{\bf k}')+\omega_0+i\omega_{k'})}\right]. \nonumber \\
\end{eqnarray}
In the above equations, $f$ and $n$ are fermionic and bosonic distribution 
functions, respectively.
The numerical integrations of (\ref{ver2}) and (\ref{cross2})
are performed by using a momentum
mesh of the Brillouin zone of $N=128\times 128$ points cheking that
increasing $N$ does not change appreciably our results.
We evaluate the real and imaginary part of the vertex and cross corrections
by making use of
two different models for the $e$-$ph$ coupling $g({\bf q})$.
In the following we report only the real part if the nonadiabatic corrections,
being the corresponding imaginary parts at least one order 
of magnitude less than the real parts

In the next section we adopt a calculation with a
momentum independent coupling $g({\bf q})=g_0$. 
This situation is therefore the one encountered
for electrons locally coupled with dispersionless phonons (Holstein model).
The effect of the momentum dependence of $g({\bf q})$ on the vertex and cross
corrections will be studied in Sec.\ref{smallq}.

\section{Momentum independent electron-phonon coupling}
\label{indep}
When we neglect the momentum dependence of the $e$-$ph$ 
coupling $g({\bf q})$ in equations (\ref{ver2}) and (\ref{cross2}), it turns out
that the vertex function $P$ depends only on the momentum transfer 
${\bf q}={\bf k}-{\bf k}'$, while the cross function $C$ depends on the
total momentum ${\bf Q}={\bf k}+{\bf k}'$.
This situation permits us to study the momentum dependence of $P$ and $C$
as a function of ${\bf q}$ and ${\bf Q}$ spanning the Brilloiun zone, respectively. 

\begin{figure}
\protect
\centerline{\psfig{figure=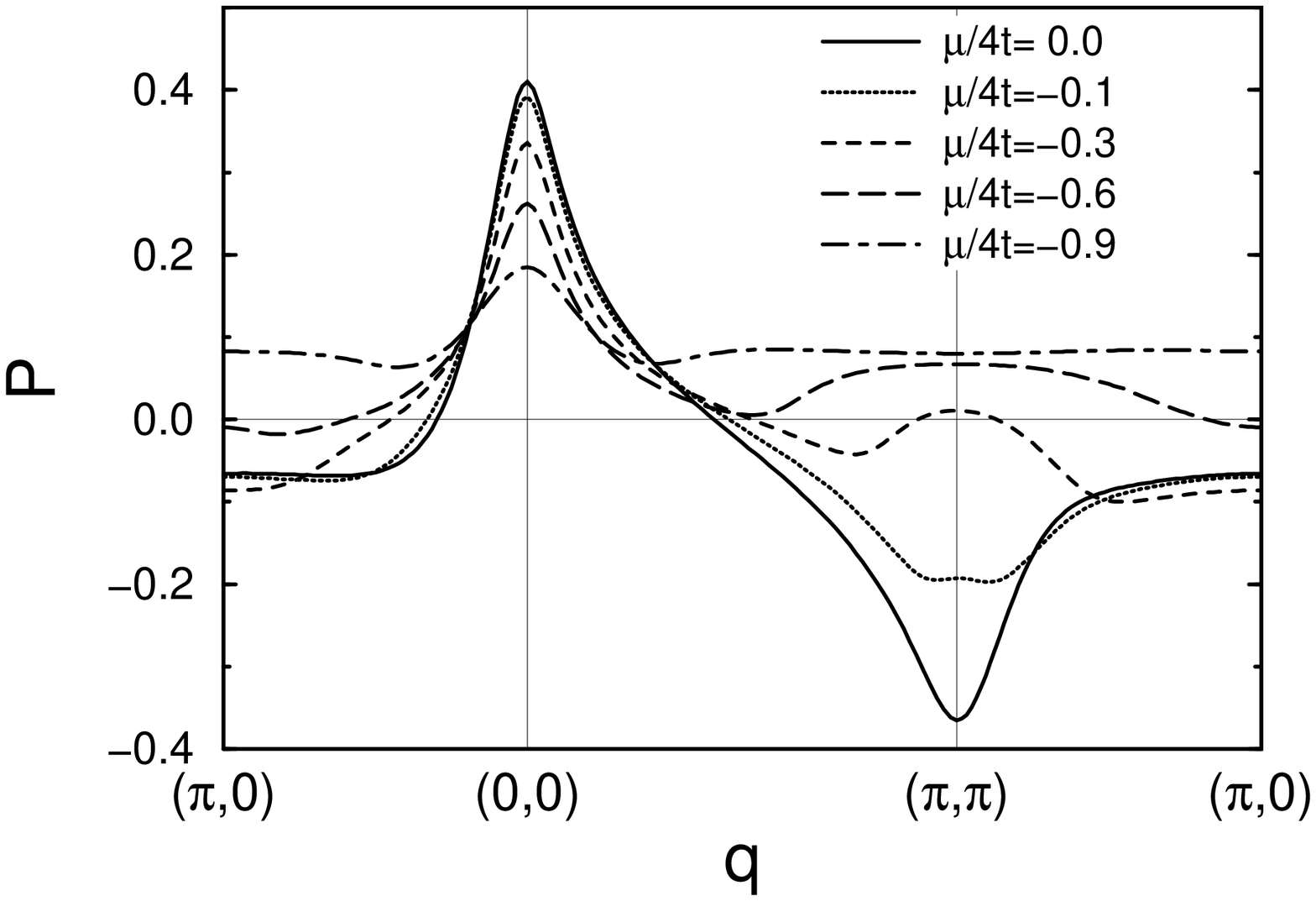,width=10cm}}
\caption{Vertex function $P$ for momentum-independent electron-phonon
coupling as a function of ${\bf q}={\bf k}-{\bf k}'$ and for
different values of the chemical potential $\mu$.
The calculation has been performed with $\omega_0/(4t)=0.2$,
$\omega_k=\pi T$ and $\omega_{k'}=\omega_0$.} 
\label{pvvsn}
\end{figure}

In Fig.\ref{pvvsn} we show the calculated vertex correction $P$ as a function
of ${\bf q}={\bf k}-{\bf k}'$ for $t'=0$ and different values of the chemical 
potential $\mu$. The calculation has been performed with a coupling
$g_0^2/(4t)=1$, a temperature small
compared to the phonon frequency $\omega_0$ ($T/\omega_0=0.01$),
and the adiabatic parameter $\omega_0/(4t)$ has been set equal to $0.2$.
The incoming and outcoming frequencies are set equal to $\omega_k=\pi T$,
that is the lowest fermionic Matsubara frequency, and 
$\omega_{k'}=\omega_0$, respectively.
At half filling ($\mu=0$), the vertex shows a strong dependence over the 
momentum transfer ${\bf q}$. A maximum positive value is found
at ${\bf q}=(0,0)$
(which corresponds to ${\bf k}={\bf k}'$), while a negative minimum is found at
${\bf q}=(\pi,\pi)$, which corresponds to the nesting vector.
Therefore this calculation confirms the qualitative analytic results obtained in previous
works,\cite{GPSprl,PSG} that is, small values of the momentum transfer give 
positive values of the vertex function at half filling.
By moving away from half filling the momentum dependence is weakened and
at the same time the region where the vertex assumes negative values is reduced. 
For $\mu/(4t)=-0.9$, near the bottom of the electronic band,
the vertex is always positive for the frequencies used in the calculation.

This behavior is remarkable, since it shows that the vertex correction is very
sensitive to the band filling and to the distance from the van Hove singularity (vHs).\cite{CP} 
This result can be understood by observing that
the strong dependence of $P$ on the electron density is due to 
many-body effects given by the Pauli exclusion principle. In fact it has been shown
that the exchange contribution of the phonon mediated electron-electron interaction
is responsible for the negative values assumed by the vertex in Matsubara
frequencies. By lowering the electron density, these many-body effects are weakened
and the vertex becomes mostly positive.\cite{GPM}
Moreover, away from half filling, the effect of the vHs is reduced and the 
${\bf q}$-dependence is reduced.

\begin{figure}
\protect
\centerline{\psfig{figure=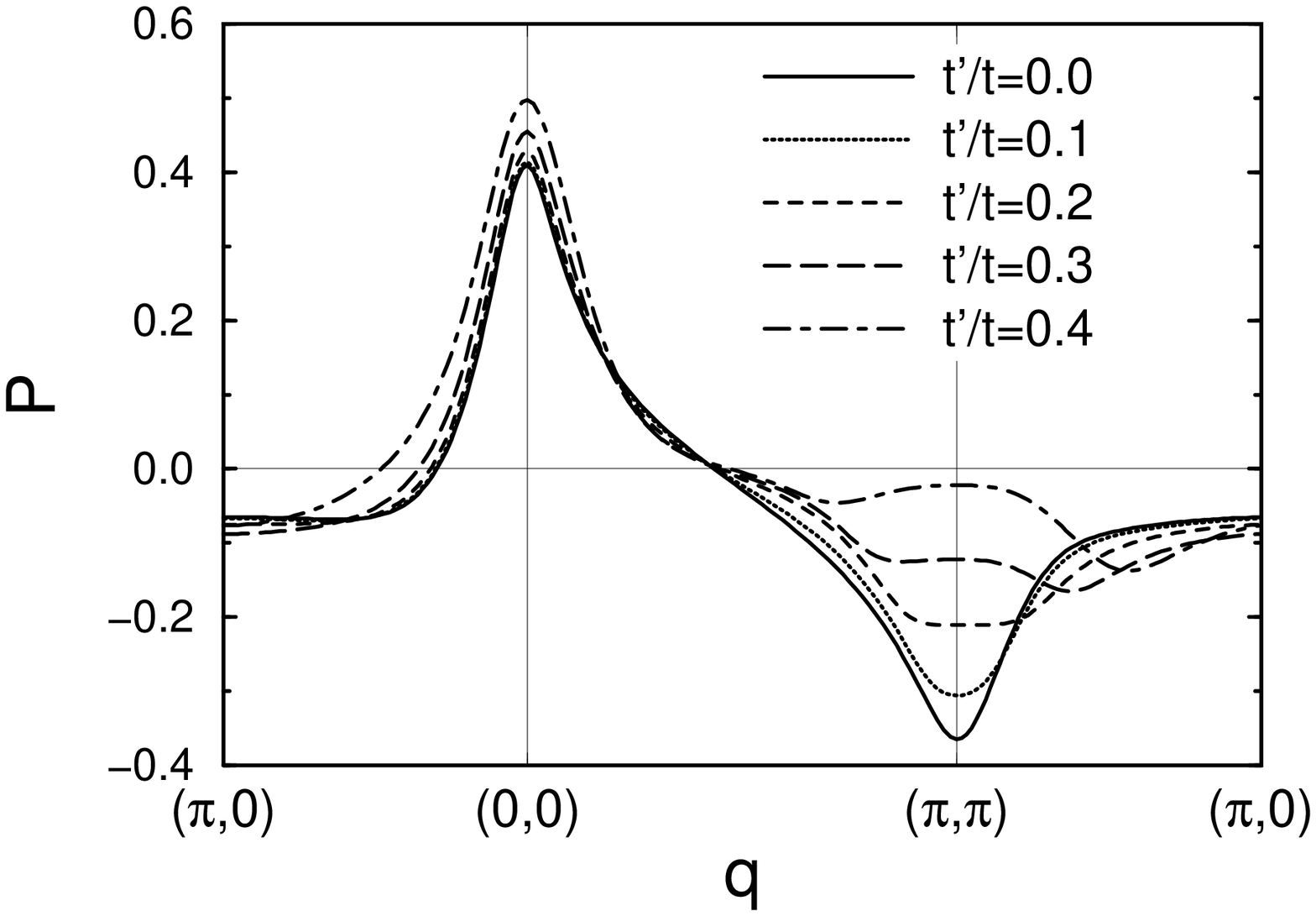,width=10cm}}
\caption{Vertex function $P$ with $\mu=-4t'$ for
different values of the next-nearest neighbor hopping
parameter $t'$. $\omega_0/(4t)=0.2$,
$\omega_k=\pi T$ and $\omega_{k'}=\omega_0$.} 
\label{pvvst}
\end{figure}

Note that in Fig.\ref{pvvsn} the strongest dependence on band filling
is in the region close to the point ${\bf q}=(\pi,\pi)$. This behavior is due to nesting
effects and disappears when nesting is destroyed. In fact, we show 
in Fig.\ref{pvvst} the behavior of the vertex when a next-nearest hopping
amplitude $t'$ is introduced. The curves for different values of $t'$ have 
been calculated by setting the chemical potential $\mu$ always coincident with the
van Hove singularity at $-4t'$, and the remaining parameters have 
the same values as in Fig.\ref{pvvsn}.
It is clear that nonzero values of $t'$
have strong effects around $(\pi,\pi)$ while for other values of ${\bf q}$
the dependence of the vertex on $t'$ is much weaker. Moreover, by
enhancing $t'$, the negative peak at ${\bf q}=(\pi,\pi)$ is weakened and for $t'/t>0.3$
is completely destroyed.

\begin{figure}
\protect
\centerline{\psfig{figure=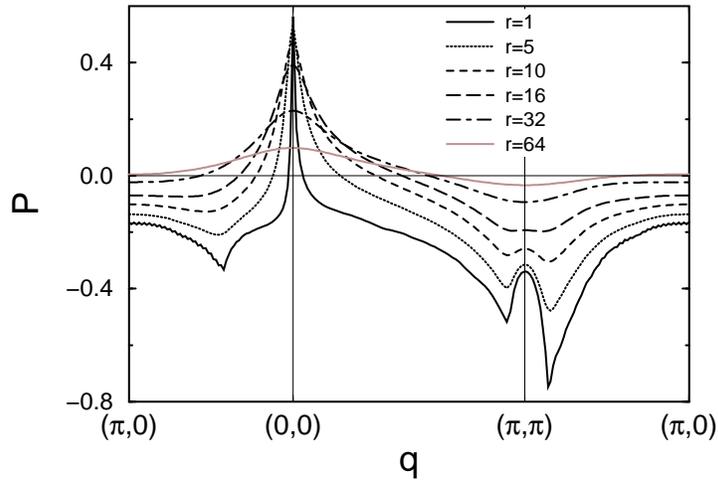,width=10cm}}
\caption{Vertex function $P$ with $\mu/(4t)=-0.1$ and $t'=0$
for different values of the outgoing frequency $\omega_{k'}$.
$\omega_0/(4t)=0.2$ and $\omega_k=\pi T$.} 
\label{pvvsw}
\end{figure}

So far, the curves of the vertex have been evaluated for frequencies fixed at 
$\omega_k=\pi T$ and $\omega_{k'}=\omega_0$. 
The same values of the incoming and outcoming frequencies
have been used in Refs.\onlinecite{GPSprl,PSG} for the analytic formula
of the critical temperature $T_c$, which is in rather
satisfactory agreement with the $T_c$ given by a fully numerical solution 
of the generalized Eliashberg equations.\cite{PSG}
However, other physical quantities besides $T_c$ are affected in a different
way by the nonadiabatic corrections. For example, the effective electronic mass
$m^*$ is mostly influenced by the vertex function calculated for
values of the exchanged frequency $\omega=\omega_k-\omega_{k'}$ 
much lower than $\omega_0$.\cite{GCP}
Therefore, it is important to analyze the role played by the exchanged frequency.
To this end, we have calculated the vertex function $P$ for $\omega_k=\pi T$
and different values of $\omega_{k'}$. 
In Fig.\ref{pvvsw} we plot $P$ for $t'=0$, $\mu/(4t)=-0.1$
and $\omega_{k'}=(2r+1)\pi T$, where the integer number $r$ runs
from $r=1$, corresponding to $\omega_{k'}\simeq 0.06\omega_0$,
to $r=64$ which gives $\omega_{k'}=4\omega_0$. For $r=16$ we
obtain $\omega_{k'}=\omega_0$ and it is therefore the same situation
as in Fig.\ref{pvvsn}.
We notice that by lowering the frequency, the vertex assumes more structure
in the momentum transfer ${\bf q}={\bf k}-{\bf k}'$ and the positive peak
at ${\bf q}=(0,0)$ becomes narrower. This behavior is consistent with the
condition $v_F|{\bf q}|<\omega$, where $v_F$ is the Fermi velocity, 
for the positivity of the vertex function.\cite{PSG}

Summarizing the above analysis, the vertex function has a strong 
momentum-frequency dependence
at half filling and it is positive for small momentum transfer.
Away from half filling, the dependence on the momentum is weakened, but
at the same time the region where the vertex is negative is reduced.
It is important to notice that in the linearized gap equation (\ref{gap1})
the vertex corrections
are always multiplied by the phonon propagator which provides a negative sign
(see Eqs.(\ref{green2},\ref{kernel1})).
Therefore, positive values of the vertex give attractive contributions
to the pairing interaction $V(k,k')$ 
and can lead to an enhancement of $T_c$.

\begin{figure}
\protect
\centerline{\psfig{figure=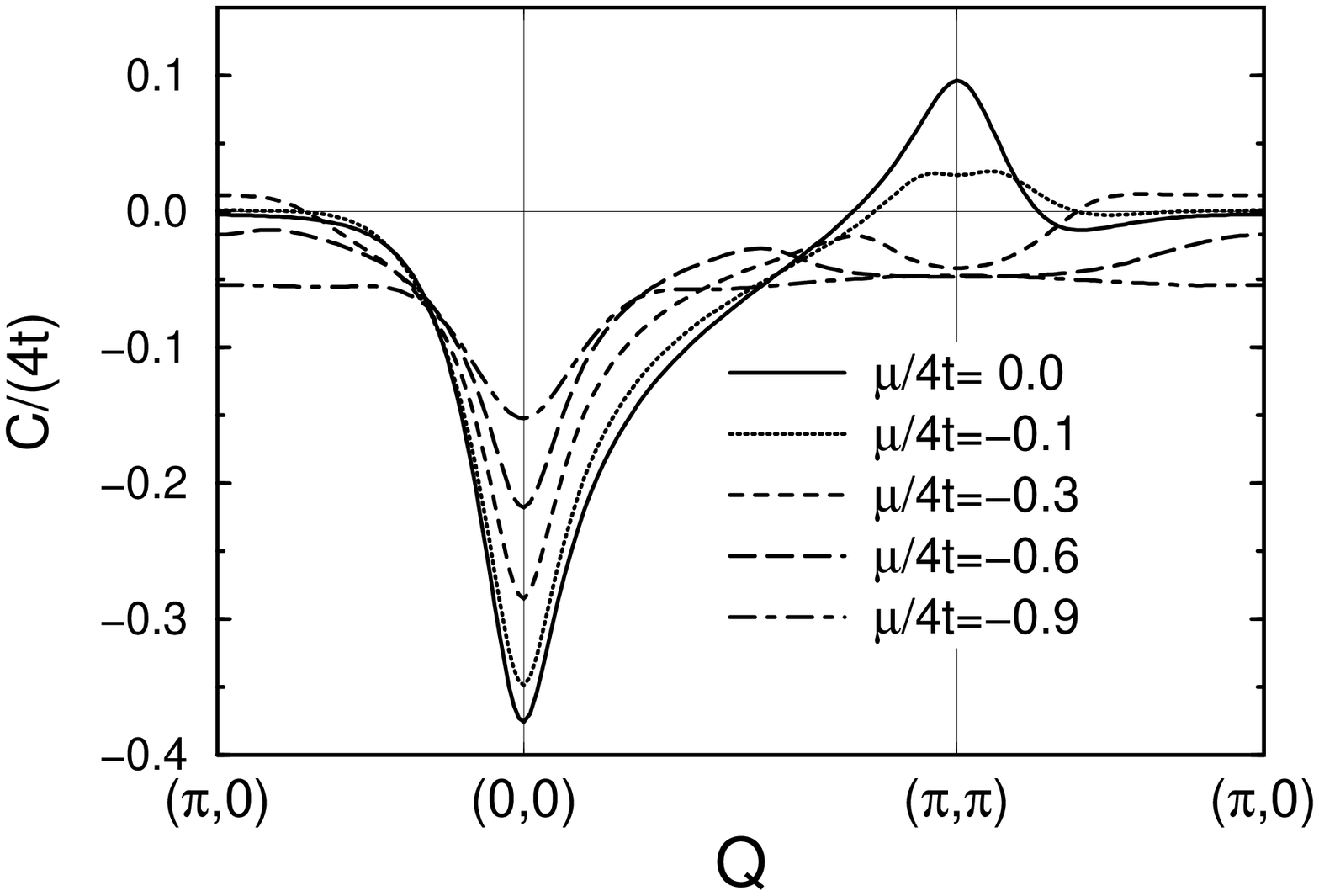,width=10cm}}
\caption{Cross function $C$ for momentum-independent electron-phonon
coupling as a function of ${\bf Q}={\bf k}+{\bf k}'$ and for
different values of the chemical potential $\mu$.
The calculation has been performed with $\omega_0/(4t)=0.2$,
$\omega_k=\pi T$ and $\omega_{k'}=\omega_0$.} 
\label{pcvsn}
\end{figure}

\begin{figure}
\protect
\centerline{\psfig{figure=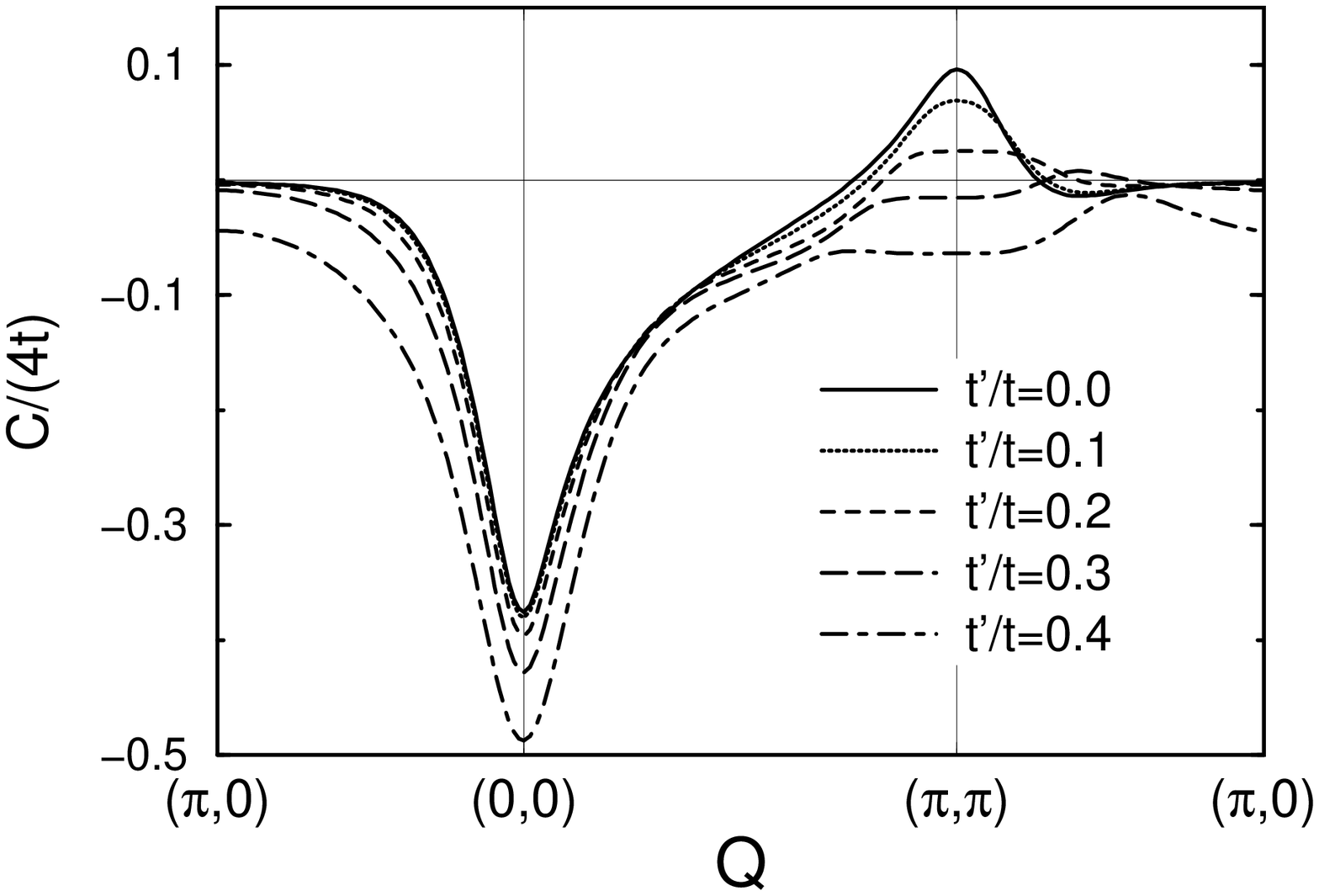,width=10cm}}
\caption{Cross function $C$ with $\mu=-4t'$ for
different values of the next-nearest neighbor hopping
parameter $t'$. $\omega_0/(4t)=0.2$,
$\omega_k=\pi T$ and $\omega_{k'}=\omega_0$.} 
\label{pcvst}
\end{figure}

The question now concerns the behavior of the other nonadiabatic correction
appearing in the kernel of the generalized Eliashberg equations,
{\it i.e.}, the cross function $C$ (\ref{cross2}).
As we have said at the beginning of this section, when the $e$-$ph$
coupling is momentum independent, the cross function depends on the
external momenta ${\bf k}$ and ${\bf k}'$ only through the total momentum
${\bf Q}={\bf k}+{\bf k}'$. This momentum dependence is quite different from 
the one relevant for the vertex correction (${\bf q}={\bf k}-{\bf k}'$).
However, the results for the cross are qualitatively the same of those for 
the vertex provided we change ${\bf q}$ in ${\bf Q}$.
In fact, we show in Fig.\ref{pcvsn} the cross correction in units of $4t$ 
as a function of ${\bf Q}={\bf k}+{\bf k}'$ for 
different fillings. As in Fig.\ref{pvvsn}, we set
$t'=0$, $\omega_0/(4t)=0.2$, $\omega_k=0$ and $\omega_{k'}=\omega_0$.
At half filling ($\mu=0$), the cross function is attractive for ${\bf Q}=(0,0)$
and repulsive at $(\pi,\pi)$. Away from half filling, the ${\bf Q}$-dependence 
is weakened and for $\mu/(4t)=-0.9$ the cross remains always attractive.
As for the vertex function, the feature at ${\bf Q}=(\pi,\pi)$ at half filling
is given by nesting effects. These effects are removed for $t'\neq 0$
as it is shown in Fig.\ref{pcvst} where the chemical potential 
is set equal to the vHs energy ($\mu=-4t'$).

From the above discussion, the situation
for the effective pairing interaction $V(k,k')$, Eq.(\ref{kernel1}), 
appearing in the linearized 
gap equation (\ref{gap1}) is the following.
The contributions coming from
the vertex depend on ${\bf q}={\bf k}-{\bf k}'$ and, near half filling, they
are attractive for ${\bf q}\simeq (0,0)$ while the cross
term is attractive for ${\bf Q}={\bf k}+{\bf k}'\simeq (0,0)$. 
The structure in the momenta of the effective interaction
is therefore nontrivial because of the differences between the vertex 
and cross dependence over momenta.

In this situation, the momentum
dependence of the pairing interaction $V(k,k')$ is better
described by employing the angular representation shown in fig.\ref{fs} and
described as follows.
We consider both the momenta
on the Fermi surface ${\bf k}={\bf k}_F$ and ${\bf k}'={\bf k}'_F$ and
we define the angle $\phi$ between these vector momenta with respect to the center
of the closed holes Fermi surface centered around the $M=(\pi,\pi)$ point of
the Brillouin zone. The momentum ${\bf k}_F$ is then fixed each times on relevant
few points and ${\bf k}'_F$ runs on the whole Fermi surface
taking into account small and large ${\bf q}$ scattering in all directions.
This angular representation is also very useful when we shall study in the next 
section the effects of the momentum-dependence of the $e$-$ph$ interaction.

\begin{figure}
\protect
\centerline{\psfig{figure=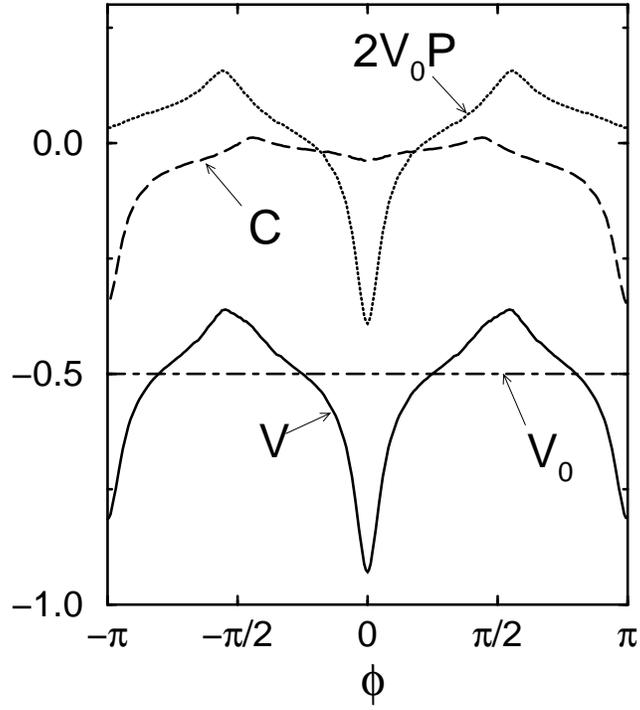,width=13cm}}
\caption{Vertex $V$, cross $C$, bare $V_0$ and
effective $V$ pairing interactions for momentum-independent 
electron-phonon coupling as a function of the
angle $\phi$ between ${\bf k}_F$ and ${\bf k}'_F$.
All data are in units of $4t$.
The calculations are performed by setting $\omega_k=\pi T$
and $\omega_{k'}=\omega_0$, so that the bare structureless
interaction $V_0$ is close to $-0.5$.} 
\label{pvpcker1}
\end{figure}

The resulting angular representation of the vertex $P$, the cross $C$ and the total
pairing interaction $V$ is shown in Fig.\ref{pvpcker1}.
The calculation has been performed with the same set of values as in
Fig.\ref{pvvsn} with the chemical potential $\mu/(4t)=0.1$, {\it i.e.}, 
slightly above
the van Hove singularity corresponding to an open electron Fermi surface
(closed holes Fermi surface centered around $M=(\pi,\pi)$).
The momentum ${\bf k}_F$ is fixed at $(k_{xF},\pi)$ near the saddle point
located at $Y=(0,\pi)$ where the local density of states
$n({\bf k})=1/|\mbox{\boldmath $\nabla$}\epsilon({\bf k})|$
$(LDOS)$ diverges.
The vertex $P$ has the maximum positive value for $\phi=0$ (${\bf q}=(0,0)$)
corresponding to an attractive contribution for $2V_0(k-k')P(k,k')$
at $\phi=0$;
the cross $C$ has the minimum negative value for $\phi=\pi$ (${\bf Q}=(0,0)$)
and a small local minimum negative value for $\phi=0$
(${\bf Q}=(2k_{xF},2\pi)$).
The total effective interaction $V$, Eq.(\ref{kernel1}), 
acquires therefore a strong momentum dependence ($\phi$-dependence)
solely due to the nonadiabatic corrections. Compared to the bare
structureless interaction $V_0$, the effective pairing $V$ is enhanced
at $\phi\simeq 0$ (due to vertex contributions) and at $\phi\simeq\pm \pi$
(due to the cross contribution). 
Note that $V$ is weakened at $\phi\simeq\pm\pi/2$ with respect to $V_0$.
These results lead to the following qualitative consideration.
If we imagine to add a repulsive interaction $U$ chosen in
such a way that, for frequencies of order $\omega_0$, 
$U+V_0\simeq 0$, the inclusion of the nonadiabatic 
corrections leads to the total effective interaction
$U+V\simeq 2V_0P+C$ which has a momentum dependence consistent
with a $d$-wave like symmetry of the gap parameter.

The result of this section is that the vertex and cross corrections have structures
which give attractive interaction at different point of the Brillouin zone. It has to
be stressed that, although we start with a momentum independent $e$-$ph$
interaction $g({\bf q})=g$, the nonadiabatic corrections give rise to a strong 
momentum dependence of the kernel for the superconducting instability.
In view of the present analysis, it appears therefore that in 
calculating the critical temperature $T_c$ beyond Migdal's limit one has to
deal with the strong frequency-momentum dependence of the nonadiabatic corrections.
For example, In Ref.\onlinecite{zieli} the authors employ an approximation
scheme which automatically neglects the momentum-frequency interplay
of the vertex and cross corrections and this leads to a reduction of the
critical temperature $T_c$.

We will see in the following section, that when we consider a model for which the
$e$-$ph$ coupling has peaked structure for small momentum transfer, the
nonadiabatic corrections, and in particular the cross function, change drastically
their momentum dependence with respect to the case studied in this section.

\section{Small $q$ electron-phonon coupling}
\label{smallq}

We consider in this section the momentum dependence of the $e$-$ph$
coupling $g({\bf q})$ using again the tight-binding electron
dispersion (\ref{disp}) and in particular we study the small $q$ limit
corresponding to almost forward scattering or long wavelenght phonons.
Attractive effective couplings peaked at small
momenta are naturally obtained in models with strong local repulsive term
and short range attractive terms such as Hubbard-Holstein model,
$t-J$ models and three bands Hubbard model.\cite{GC,CDG,ZK}
All these models show phase separation instability signalled by divergent
scattering amplitude at zero momenta; near phase separation the large
charge fluctuations can mediate the $e-e$ interactions and give rise to
pairing in the Cooper channel with $d$-wave or strongly anisotropic
$s$-wave symmetry of the gap parameter.\cite{Perali}
An $e$-$ph$ coupling peaked at small momentum transfer 
can also be the consequence of weak screening
effects in a system with small charge carrier density.
\cite{Abrikosov,weger,Sigmund}

In our analysis we consider the following form for the 
$e$-$ph$ coupling $g({\bf q})$

\begin{equation}
\label{gsq}
g({\bf q})^2=\frac{\gamma(q_c)}{q_c^2+\omega({\bf q})^2} ,
\end{equation}
where

\begin{equation}
\label{omega}
\omega({\bf q})^2=2\left[2-\cos(q_x)-\cos(q_y)\right].
\end{equation}
This coupling has the periodicity of the lattice and,
for small values of the momentum transfer ${\bf q}$, 
$\omega({\bf q})^2\simeq |{\bf q}|^2$ so that $g({\bf q})^2$
reduces to a lorentzian form. In this way it is
easy to understand the role played by the parameter $q_c$:
it provides a momentum cutoff in the $e$-$ph$ interaction
in such a way that for $|{\bf q}|>q_c$ the coupling is
depressed. Therefore the coupling (\ref{gsq}) and (\ref{omega})
reproduces qualitatively the momentum dependence 
of the $e$-$ph$ interaction
in the presence of strong electronic correlation.

\begin{figure}
\protect
\centerline{\psfig{figure=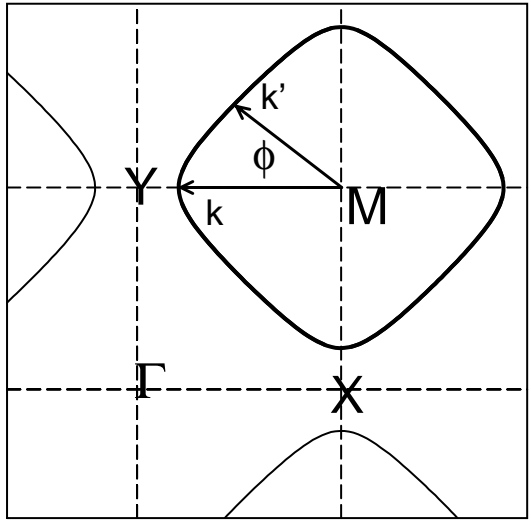,width=10cm}}
\caption{Angular $\phi$-representation.
The momentum ${\bf k}$ is fixed at one point of the Fermi surface
and the angle $\phi$ measures the position of ${\bf k}'$ with
respect to ${\bf k}$.} 
\label{fs}
\end{figure}

In order to compare the results on the nonadiabatic 
corrections calculated by using the model (\ref{gsq}) 
and (\ref{omega})
with the structureless $e$-$ph$ coupling case of the 
previous section, we introduce
the quantity $\gamma(q_c)$ defined in order
to satisfy the following normalization condition:

\begin{equation}
\label{norma}
\langle g({\bf k}-{\bf k}')^2\rangle_{{\bf k}'}=
\frac{\int^{+\pi}_{-\pi} d\phi_{{\bf k}'}n({\bf k}')
g({\bf k}-{\bf k}')^2}
{\int^{+\pi}_{-\pi} d\phi_{{\bf k}'}n({\bf k}')} 
=g^2_0,
\end{equation}
where $n({\bf k}')=1/|\mbox{\boldmath $\nabla$}\epsilon({\bf k}')|$ 
is the $LDOS$ and $g_0^2$ is independent of $q_c$.
Such a normalization ensures that even if the modulation of 
$g({\bf q})^2$ over the momentum transfer is governed by $q_c$,
its ${\bf k}'$-average over the Fermi surface remains equal to $g_0^2$.
From the above condition, $g_0^2$, and so $\gamma(q_c)$,
depends implicitly on the incoming electron momentum ${\bf k}$.
However in the following analysis we keep ${\bf k}$ fixed
on the $Y$ point of the Fermi surface (see Fig.\ref{fs})
and we set $g_0^2/(4t)=1$, as in the previous section.

From equations (\ref{ver2}) and (\ref{cross2}), it is clear that the
structure in ${\bf q}$ of the $e$-$ph$ coupling $g({\bf q})$
leads to a momentum dependence of the nonadiabatic corrections which 
cannot be described solely in terms of ${\bf q}={\bf k}-{\bf k}'$
for the vertex $P$ and ${\bf Q}={\bf k}+{\bf k}'$ for the
cross $C$.
For this reason, we employ the
angular representation introduced in the previous section.
This representation is also very useful for small $q$
( and so small $\phi$ ) interactions which couples only nearby states for a
given point ${\bf k}$ in momentum space driving the electronic systems in the
momentum decoupling $(MD)$ regime.\cite{Sigmund,VPCP} 
When $MD$ is achieved for small values
of the cutoff $q_c$ the anisotropy of several physical quantity
such as the gap parameter, the tunneling conductance
and the photoemission lineshape are driven by the $LDOS$.\cite{VPCP}
It is evident that $MD$ have to be relevant also for the ${\bf k}_F$
and ${\bf k}'_F$ dependence of the vertex and cross corrections in the
small $q$ limit. 

\begin{figure}
\protect
\centerline{\psfig{figure=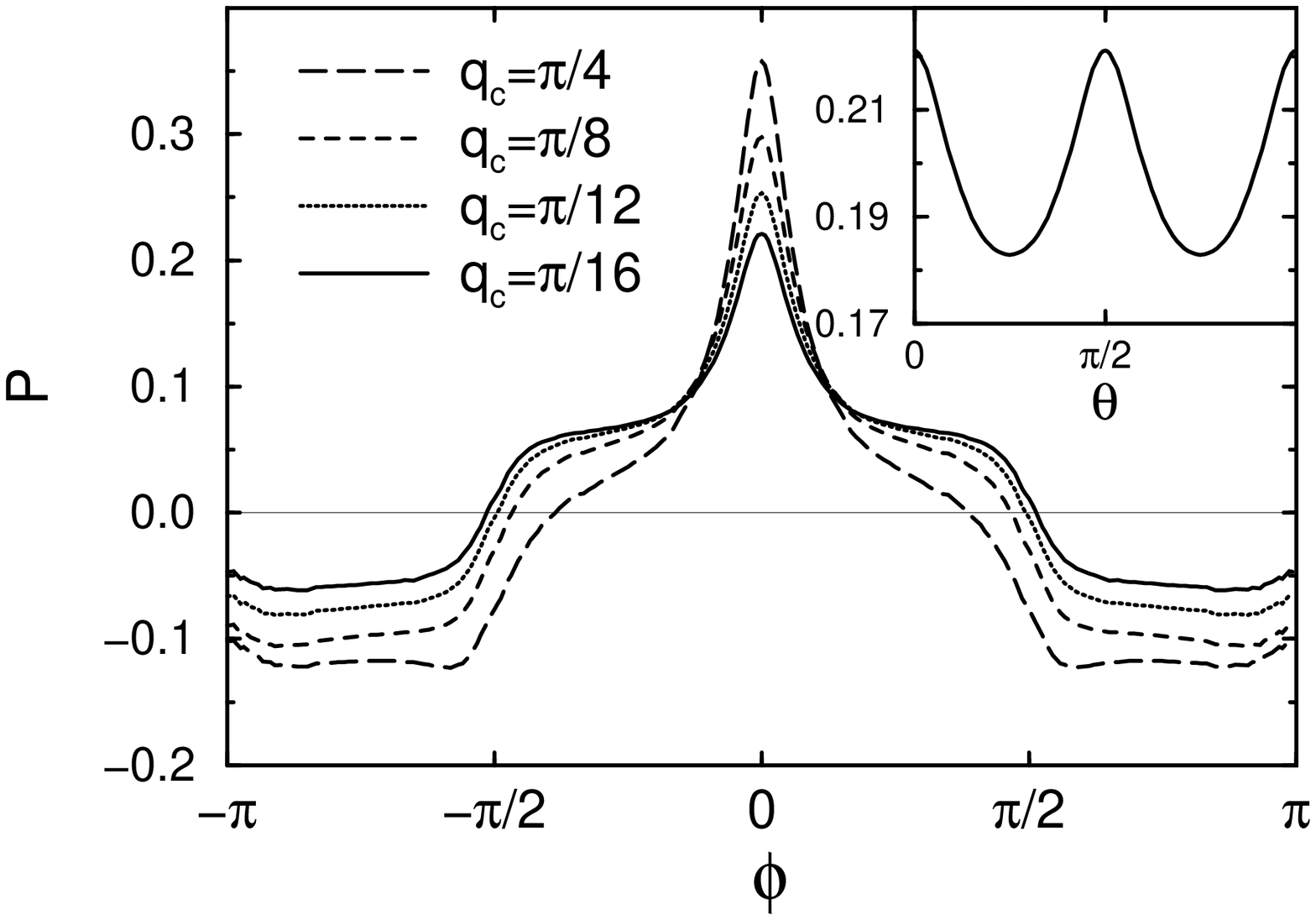,width=10cm}}
\caption{Vertex function $P$ for momentum-dependent 
electron-phonon coupling 
in the angular representation (see text).
$\omega_0/(4t)=0.2$,
$\omega_k=\pi T$ and $\omega_{k'}=\omega_0$.} 
\label{pvang1}
\end{figure}

In Fig.\ref{pvang1} we show the calculated vertex $P$ correction
as a function of the angle $\phi$ for the optical phonon with
momentum dependent coupling $g({\bf q})$ for different values of the 
coupling cutoff $q_c =\pi/16;\pi/12;\pi/8;\pi/4$.
The incoming electron momentum ${\bf k}$ has been chosen as in 
Fig.\ref{fs} obtained with $t'=0$ and $\mu/(4t)=0.1$. 
The incoming and outcoming frequencies has been set equal to
$\omega_k=\pi T$ and $\omega_{k'}=\omega_0$, respectively,
and as in the previous section the adiabatic parameter is 
$\omega_0/(4t)=0.2$. 

For all values of the momentum cutoff $q_c$, the vertex behaves
in a way qualitatively similar to the one show in Fig.\ref{pvpcker1}
for the structureless $e$-$ph$ case: $P$ is positive for small values
of $\phi$ and negative for large values.
However, in the total pairing interaction $V(k,k')$ appearing in 
Eq.(\ref{kernel1}),
the vertex $P(k,k')$ is multiplied by $g({\bf k}-{\bf k}')^2$
which is peaked at small values of ${\bf k}-{\bf k}'$
for small momentum cutoff $q_c$.
As a consequence, $g({\bf k}-{\bf k}')^2$ selects mainly the positive
contributions of the vertex correction $P$.
In the inset of Fig.\ref{pvang1} we show the vertex correction for 
$q_c=\pi/16$ evaluated at zero momentum transfer ($\phi=0$) 
for different incoming electron running on the 
Fermi surface defined by the angle $\theta$;
the maximum of the vertex has a weak dependence on $\theta$ and its anisotropy
is driven by the $LDOS$. The effect is not strong but it is on the right
direction with respect to the $MD$, giving the maximum correction near the
saddle points where the $LDOS$ is high and the minimum near the 
$(\pi/2,\pi/2)$ point of the Brillouin zone where the $LDOS$ is low.

\begin{figure}
\protect
\centerline{\psfig{figure=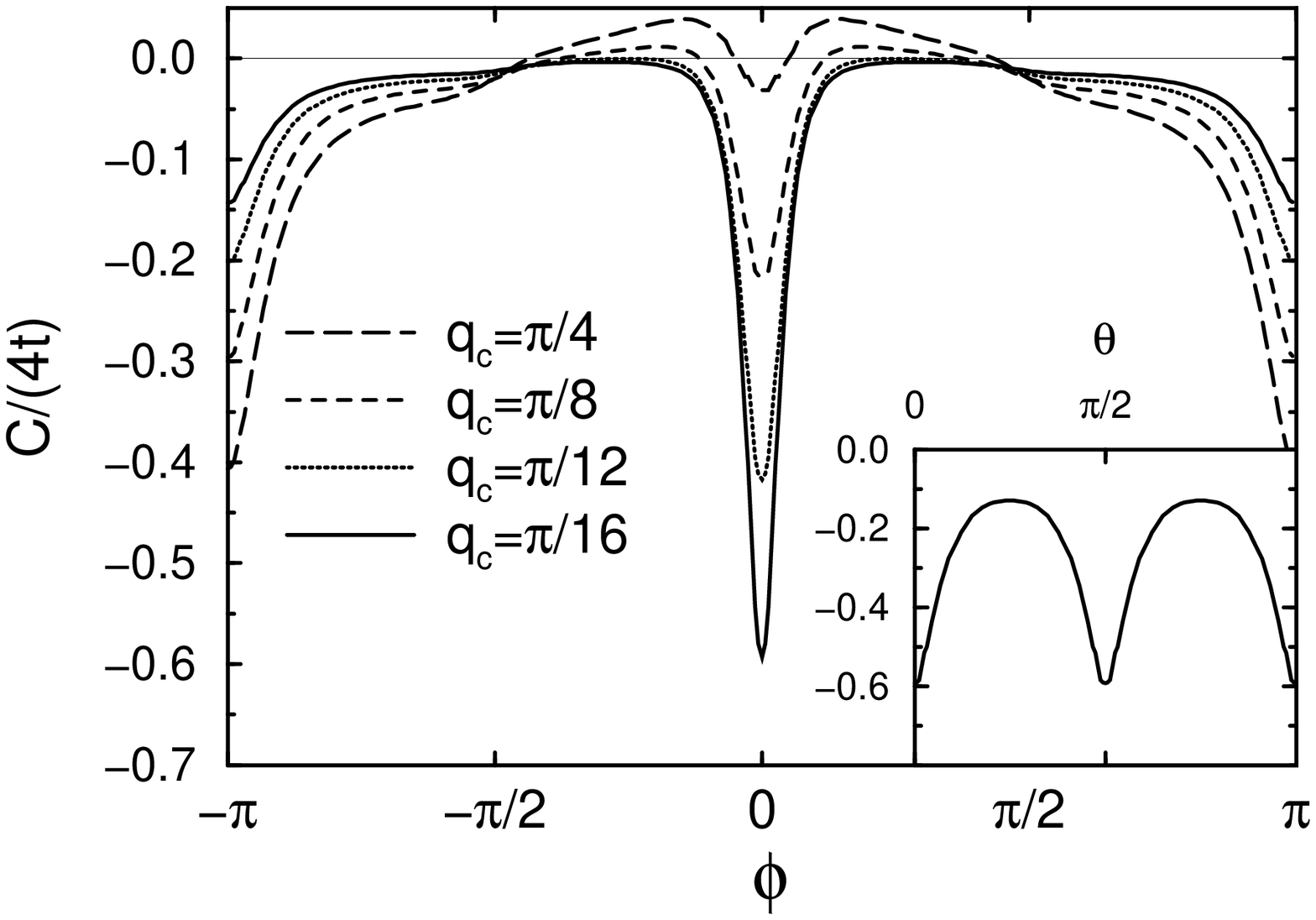,width=10cm}}
\caption{Cross function $C$ for momentum-dependent 
electron-phonon coupling 
in the angular representation (see text).
$\omega_0/(4t)=0.2$,
$\omega_k=\pi T$ and $\omega_{k'}=\omega_0$.} 
\label{pcang1}
\end{figure}

In Fig.\ref{pcang1} we show the calculated cross correction $C$ 
for different
values of the coupling cutoff $q_c =\pi/16;\pi/12;\pi/8;\pi/4$.
The angular dependence $\phi$ of $C$ is strongly sensitive to the
cutoff value $q_c$.
For $q_c=\pi/4$ we recover a situation similar to the one obtained for
momentum independent coupling: the relevant attractive contribution
arises for large $\phi$ values (${\bf Q}\simeq 0$).
When the $e$-$ph$ coupling becomes peaked at small ${\bf q}$ the
$\phi$ dependence of  $C$ is completely different: the relevant attractive
region moves from large to small $\phi$ values further increasing,
together with the vertex, the total attractive correction
for small momentum transfer.
In the inset of Fig.\ref{pcang1} we show the cross correction for
$q_c=\pi/16$ evaluated at ${\bf Q}=2k_{F}$ $(\phi=0)$ as a function
of $\theta$. 
The small ${\bf q}$ coupling gives rise to strong anisotropy 
in the cross correction
and near the saddle points is four times bigger than near the 
$(\pi/2,\pi/2)$ point. Then the cross correction increases the tendency to
$MD$ already given by the momentum dependent small ${\bf q}$ coupling.

\begin{figure}
\protect
\centerline{\psfig{figure=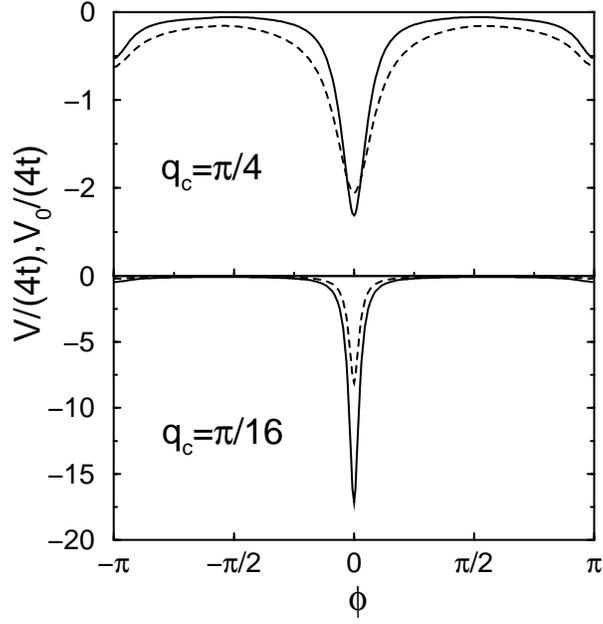,width=10cm}}
\caption{Total pairing $V$ (solid lines) 
compared with the bare one $V_0$ (dashed lines) in the
angular representation.
$\omega_0/(4t)=0.2$,
$\omega_k=\pi T$ and $\omega_{k'}=\omega_0$.} 
\label{kerang1}
\end{figure}

\begin{figure}
\protect
\centerline{\psfig{figure=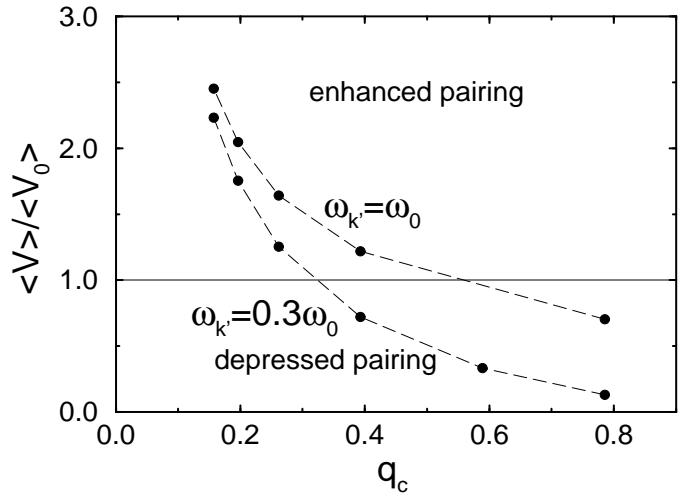,width=10cm}}
\caption{Averaged total pairing normalized to the
averaged bare interaction as a function of the momentum
cutoff $q_c$ for two different outcoming frequencies. } 
\label{kerqc}
\end{figure}

Finally we discuss the effect of small ${\bf q}$ $e$-$ph$
coupling on the effective pairing interaction $V(k,k')$, 
Eq.(\ref{kernel1}), keeping in mind the above results.
In Fig.\ref{kerang1}, we report the pairing $V(k,k')$ in the
angular $\phi$ representation (solid lines) for two different
values of the momentum cutoff $q_c$ compared with the bare pairing
$V_0(k-k')$ (dashed lines).
For $q_c=\pi/4$ (top panel) we find a slight pairing enhancing
for small $\phi$ and a weak depression for large $\phi$; 
for $q_c=\pi/16$ (bottom panel) we find a large amplification 
at small $\phi$ leading to an overall enhancement of the pairing.
Therefore, by lowering the value of the momentum cutoff $q_c$
we expect that the nonadiabatic corrections
enhances the averaged effective pairing.
This situation is confirmed in Fig.\ref{kerqc} 
where we plot the momentum average of $V(k,k')$ in units
of the averaged bare interaction $V_0(k-k')$ as a function
of the cutoff $q_c$ and for two different values of
the outcoming frequency $\omega_{k'}$. 
The averages are performed as described in
Eq.(\ref{norma}), ${\bf k}$ being fixed on the $Y$ point of the
Fermi surface. We find that, for frequency $\omega_{k'}=\omega_0$,
$\langle V\rangle/\langle V_0\rangle$
is greater than unity for $q_c<\sim 0.5$ and that for a lower
frequency ($\omega_{k'}=0.3\omega_0$) 
the enhancement is found for a smaller value
of $q_c$. 

Summarizing, in this section we have shown how the 
$e$-$ph$ coupling peaked at small momenta selects the attractive
contribution of the nonadiabatic corrections giving an
overall amplification of the pairing interaction.

\section{Conclusions}
\label{conc}

We have addressed the calculation of the vertex $P$ and the cross $C$ 
corrections to the
Migdal-Eliashberg equations for specific band structures by
numerical integration over the momentum space. 
We have considered a two-dimensional tight-binding model for the electron
dispersion including near and next-near-neighbors hopping $(t,t')$ and an 
$e$-$ph$ coupling in the nonadiabatic regime with an optical phonon 
of frequency $\omega_0/(4t)\simeq 0.2$.
Starting from a momentum independent $e$-$ph$ coupling the $P$ and $C$
corrections provide a strong momentum and frequency dependence
to the effective pairing. We find that the strenght of the pairing
is enhanced for incoming electron momenta close to the outcoming
ones due to the $P$ corrections and for almost opposite
momenta due to the $C$ correction.

Considering a momentum dependent $e$-$ph$ coupling, we find that
small ${\bf q}$ peaked interactions select the attractive contribution
of the vertex and cross terms leaving out the repulsive part,
leading to an overall amplification of the total pairing.

Our main result is that the nonadiabatic vertex and cross corrections
beyond Migdal's theorem lead to attractive contributions to the
electron-electron interaction in the Cooper channel.
These effects can provide a strong amplification of the
superconducting critical temperature $T_c$.
Moreover, the strong momentum dependence of the effective
pairing induced by the nonadiabatic corrections can lead to
strong anisotropy in the gap parameter and eventually to
$d-$wave symmetry when a residual electron-electron repulsion
is acting in the system.

\acknowledgments
C. G. acknowledges the support of a I.N.F.M. PRA project.

\end{document}